# Wave formation process upon explosive welding: electron microscopic observations and imitating experiments

B.A. Greenberg[1,a], M.A. Ivanov[b], A.V. Inozemtsev[a], M.S. Pushkin[a,c], A.M. Patselov[a], O.A. Elkina[a], S.V. Kuzmin[d], V.I. Lysak[d]

[a]*M.N. Miheev Institute of Metal Physics, Ural Branch, Russian Academy of Sciences, S. Kovalevskoi str. 18, Ekaterinburg, 620990, Russia*
[b]*Kurdyumov Institute of Metal Physics, National Academy of Sciences of Ukraine, Vernadskogo blvd. 36, Kiev, 03680, Ukraine*
[c]*The Ural Federal University named after the first President of Russia B.N.Yeltsin, Mira str. 19, Ekaterinburg, 620002, Russia*
[d]*Volgograd State Technical University, Lenin Av. 28, Volgograd, 400005, Russia*

ABSTRACT

A sequence of the interface transition states was investigated as the explosive welding mode was intensified. A transition state has been found, during which cusps, even though they are solid-phase, look like splashes on water. Images of the splashes for Cu-Ta, Al-Ta, Cu-Ti, Mg-Ti joints have been obtained. The processes of self-organization of the splashes, first into a quasi-wave surface and then - into a fairly perfect wavy surface, have been revealed. It has been found that the quasi-wave interface relief is similar to that of the steel bullet which was observed after a collision with a target, at the famous Abrahamson's experiments. Simulations have been conducted to determine possible ways of relaxation of a nonequilibrium structure having an excess area. Heterogeneity and the interface bending of one of the contacting materials have been detected. They imitate the heterogeneity on the interface arising upon the explosive welding.

## 1. Introduction

It is well known that the explosive welding is a highly intense and rapid action on the material which allows producing a large number of joints, including those that could not be manufactured in other manner. The existence of the so-called "weldability window" (in coordinates of "collision angle - collision velocity") is important for describing the necessary conditions for the formation of a strength joint [1-3].

---

[1] bella@imp.uran.ru

Using of welding mode far from the lower boundary of the weldability window the formation of a wavy interface occurs which is one of the most characteristic and most prominent phenomena. It distinguishes the explosive welding process from the other ways of joining of materials. For some time it was generally accepted that it was the wave formation that was responsible for the quality of the joint. But it turned out that in many cases (titanium - steel, aluminum - steel) the best quality welding is achieved without formation of waves. As noted in [3], the character and parameters of the waves that form on the interface depend on a significant number of factors, including the velocity and angles of the impact, as well as on the characteristics of the contacting materials. The wave formation is primarily a process of irreversible deformation on the collision surface. Therefore, it is obvious that the pressure, developed in the impact, should at least exceed the dynamic yield stress of the colliding materials. Numerous attempts have been made to explain the mechanism of the wave formation [1-3], but still to this day its nature is not fully understood and there is no commonly accepted theory for this phenomenon. However, the fact of occurrence of the periodic relief of the interface remains amazing. It is worth noting here, that for the identification of the wave formation criteria it would be very useful to conduct a thorough investigation of such impact modes and occurring structures when the wave formation is unstable, i.e., it can disappear when a small change of a parameter of the impact occurs.

One of the first attempt to explain the nature of the wave formation belongs to Abrahamson [4]. As a result of experimental data analysis for the impact of a steel bullet on a target, it was demonstrated that in the area of the impact point very high pressures arise. The paper [4] focuses on the experimental fact that the waves occur at a given velocity of collision of a flat-nosed steel bullet striking a thin lead target only when the angle of impact more than a certain critical one. As shown in [1] an attempt was made to simulate the process of wave formation in non-metallic materials: a stream of water was made to fall obliquely on a viscous substance positioned at the bottom of a slowly moving tray. It was possible to obtain a periodic deformation of the surface, while filming allowed to describe the process of wave formation.

Other works [5-7], in some cases antagonistic and in others - similar to Abrahamson's model, also used hydrodynamic models. At first glance, applicability of such models to solid crystalline bodies that remain intact in the process of explosive welding (except for some molten regions) does not seem reasonable enough. If we assume that the wave formation occurs under the influence of cumulative jets of gas before the contact of materials, it remains unclear how the congruent joining of the opposite surfaces occurs.

In this paper we will attempt to explain the mechanism of wave formation during the explosive welding based on the analysis of plastic deformation of materials in their contact area. It will be taken into account the fact that due to the strong impact, a significant deflection of materials occurs resulting in a significant increase of the area of the contact followed by an irreversible plastic deformation. Further, after an external impact the material in general returns in its initial state while the excess contact surface should, on the one hand, remain present, but on the other hand, it should return to its original state. It is this relaxation that can cause the occurrence of the wavelike character of the interface. To test this assumption a series of simulations for a number of contacting materials will be carried out. It is their joint deflection, together with their returning to the original state, that is expected to show the mechanism responsible for the relaxation of the occurring excessive contact area.

Also, attention might be paid to the fact that the existing models of wave formation do not utilize experimental data on the evolution of the interface with the intensification of welding mode. Therefore, before proceeding to a presentation of simulation experiments, in this paper we will summarize the results of electron microscopic analysis of the interface for such joints as Cu-Ta, Al-Ta, Cu-Ti, Mg-Ti. The transitional states between the plane and wavy shapes of interfaces including the splashes which occur at a low intensity of the explosive impact will be described. Further, a self-organization of splashes, first into a quasi-wave surface and then into a fairly perfect wavy surface, will be considered as the explosive welding mode is intensified.

## 2. Materials and Joints

The explosive welding has been carried out by Volgograd State Technical University, OJSC Ural Plant of Chemical Engineering, Ekaterinburg. A parallel arrangement of plates was used. An explosive substance charge was placed on the top (cladding) plate. The following pairs of materials were selected as the starting: copper - tantalum and magnesium - titanium as they do not have mutual solubility; aluminum - tantalum and copper - titanium as they have fairly high mutual solubility.

The explosive welding parameters in coordinates of "collision angle $\gamma$ - collision velocity $V_c$" are given in [8, 9] for copper - tantalum joints, in [10, 11] for aluminum - tantalum joints, in [12] for magnesium - titanium joints, in [13] for copper - titanium joints.

Different welding modes have been used for various joints: lower than the lower boundary (LB) of the "weldability window", near the LB, upper the LB. The lower boundary is important both for practical calculations of welding modes and for understanding of the processes that determine the possibility of formation of a welded joint. The welding modes near the LB are characterized by minimal collision velocities that provide a strength joint formation [14].

Metallographic analysis was performed by means of the Epiquant optical microscope equipped with a computing system SIAMS. Experimental studies of the microstructure were carried out using the transmission electron microscopes JEM200CX and SM-30 Super Twin, scanning electron microscopes QUANTA 200 FEI Company and Quanta 600 and Fashione 1010 ION MILL ion gun was used for the preparation of foils. In order to study the surface roughness of the starting materials, Zygo NewView 7300 optical profilometer was used.

A parallel study of joints featuring both plain and wavy interfaces is very uncommon for the same pair of dissimilar materials. However, this comparison has proved to be expedient for subsequent analysis.

## 3. Results and Discussions

### 3.1. Sequence of the interface transition states

During a study of a large number of joints at various welding modes, it was found, first of all, that the interface was not smooth. It contains two kinds of inhomogeneities: cusps and zones of local melting [15]. A detailed discussion of these zones is considered in [16]. Their formation occurs in the following scenario: an in-flight dispersion of particles of the refractory phase, their deceleration at a barrier, friction, local heating, local melting of the low-melting phase. The role of the local melted zones may be twofold: a threat to the continuity of the joint, or point-to-point gluing.

Cusps were first discovered for the titanium - orthorhombic titanium aluminide joints [17, 18]. A numerous observations of the cusps have been obtained for all the joints studied, both for the flat and wavy interfaces, regardless of the mutual solubility of the initial elements [19]. Cusps occur as a result of diffusionless ejection of one welding metal into another, taking place at the time of the explosion. The geometry of the cusps indicates that they are formed by a metal that has the highest hardness for the investigated pair. As SEM analysis of the chemical composition shows, the cusps do not contain the second metal regardless of the character of their mutual solubility. In fact, in the case of the cusps, an interpenetration of the materials is supposed rather than mixing. The absence of the mixing inside the cusps is due to the fact that the diffusion in the solid phase is practically excluded because of the transience of collision. If the interface was smooth, there would have occurred problems with adhesion, and either the reconstruction of the metallic bond or transport of the point defects would have been required. But the cusps solve the surface adhesion problem, they act like "wedges" bonding the contacting metals.

The next procedure has been applied to define the relief of the interface near the LB. The low-melting phase has been etched away, and then SEM images of the surface of the refractory phase have been obtained at different angles relative to the electron beam. Dozens of micrographs have been obtained at different characteristic angles, i.e., at different positions. Only the key micrographs will be presented here.

The bonding of plates below the LB does not occur generally. The relief of the tantalum surface has a form shown in Fig. 1a for copper-tantalum joint. The isolated cusps are clearly visible. Also, a set of SEM images obtained at different angles for the same large cusp is shown here (inside the white frame).

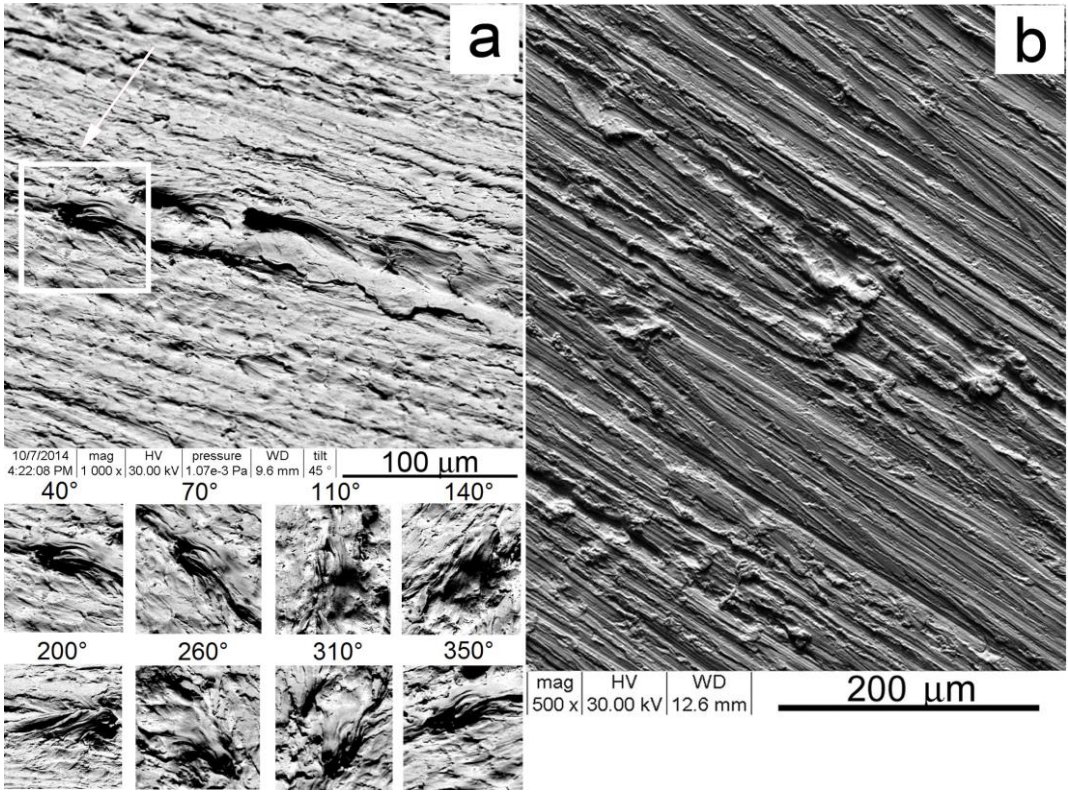

**Fig. 1.** Copper-tantalum joint (a) and aluminum-tantalum (b) with plain interface lower than LB (SEM): (a) - tantalum interface (copper etched out); (b) - tantalum interface (aluminum etched out)

The surface of the cusp, presented in Fig. 1a, contains concave segments, hollows, cavities and the other complicated heterogeneity. To ensure the continuity of the joint, they should be filled with another material both during the explosion itself, and upon the further solidification if the melt took place. In this case, for disjointing of different materials, a fairly strong force is required for a physical "destruction" of the cusps. This means that purely topological reasons are responsible for the tensile strength. In other words, the cusps may provide the topological relationship between the contacting surfaces. In this case, below LB, the amount of cusps is apparently insufficient to ensure weldability. Fig. 1b shows a SEM image of the tantalum interface for the tantalum-aluminum joint below LB.

Fig. 2a shows the image of the tantalum interface for copper - tantalum joint at LB. Even though the interface is considered to be plain, the cusps look like wave splashes on the water. This similarity is surprising, considering that the cusps have been formed by means of the solid phase that was not subjected to melting. A comparison between Figs. 1a and 2a shows that the transition from the area below the LB to the area close to it is accompanied with a dramatic change of the interface relief: separate disconnected cusps after such a relatively low intensification of the welding mode are replaced by regularly distributed splashes. Splashes have been also observed in the aluminum-tantalum (Fig. 2b), copper-titanium (Fig. 2c) and magnesium-titanium joints (Fig. 2d).

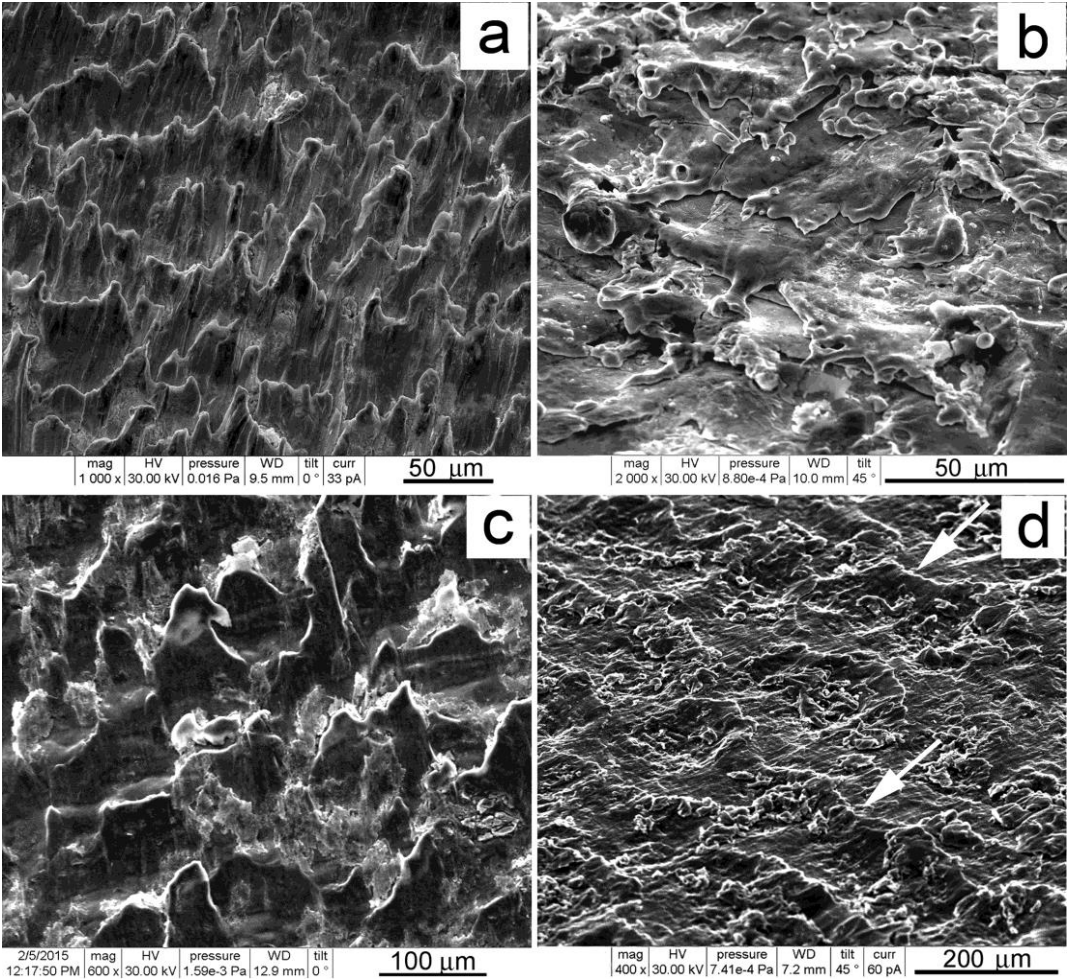

**Fig.2.** Splashes on a plain interface near LB for joints; copper-tantalum (a), aluminum-tantalum (b), copper-titanium (c), magnesium-titanium (d), (low-melting-point metall etched out): (a), (b) - tantalum interface, (c), (d) - titanium interface

Note should be taken of a significant feature of the images of the cusps: a self-similarity of its constituent elements. It is the self-similarity of cusps that was one of the factors that initiated the fractal description of the interface structure. This approach is described in detail in [16].

For different joints pictured on Fig. 2, a group of splashes is seen pressed to each other (indicated by arrows in Fig. 2d). As the welding mode is further intensified, a little above the LB, groups of cusps further arrange into rows. The splashes are associated into groups above the LB, where waves should have been formed already. This fact demonstrates that there is some affinity between the two processes. Being forerunners of the waves, the splashes accompany their occurrence from the low boundary to the center of the "weldability window". And finally, a wavy interface of copper-tantalum is visible in Figs. 3a, b but it is very inhomogeneous: there are different wave lengths and amplitudes in various areas. This interface can be referred to as "quasi-wavy". Most of all, the quasi-wavy interface resembles a "patchwork quilt" consisting of parts that have their own wave parameters and splashes along their boundaries (as indicated by arrows). A different type of quasi-wave surface is observed for copper-titanium joints. Figs. 3c, d shows the relief of the titanium interface.

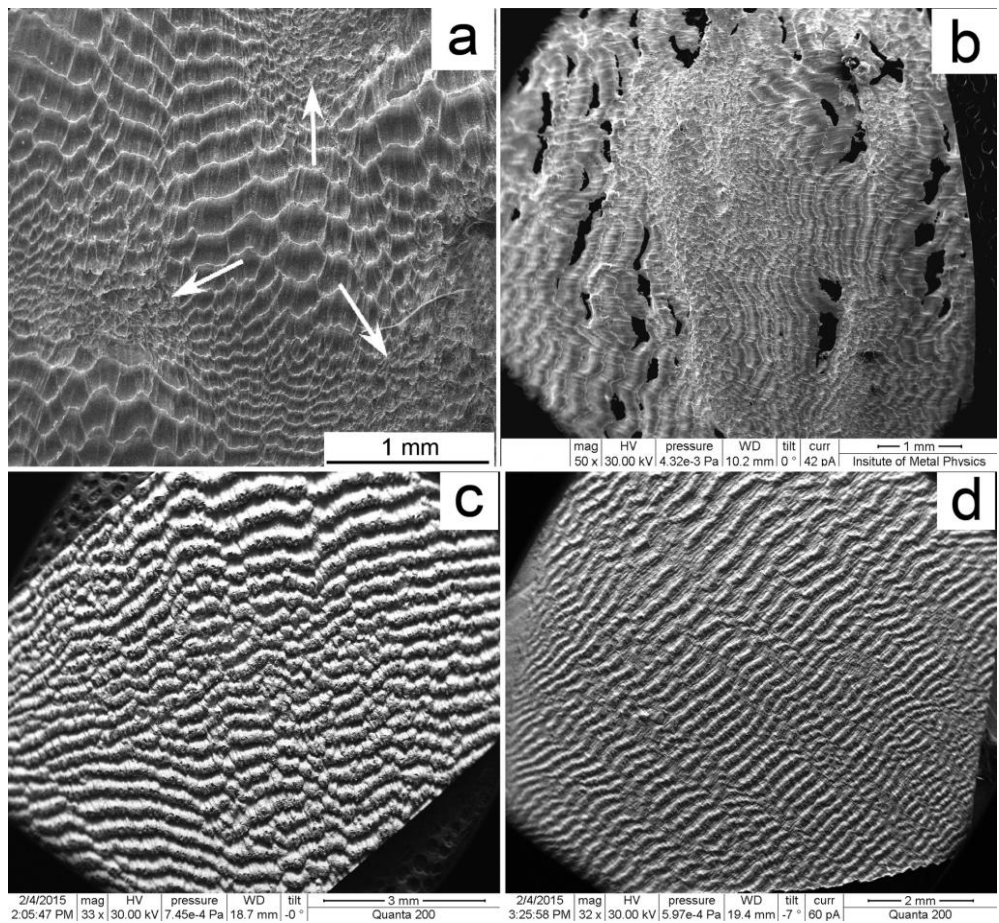

**Fig. 3.** Transition states "patchwork quilt" - type above LB: (a), (b) - copper-tantalum joints; (c), (d) - copper-titanium joints

Waves that have about the same direction of the axes, aggregate into long bands. They are separated by narrow bands where the periodic structure fails. This quasi-wave surface resembles a "patchwork quilt" consisting of the alternating bands. However, the quasi-wave surface with a random distribution of the patches which is observed for copper-tantalum joints (Fig. 3a, b) has not been found for copper-titanium joints.

In those cases, when more or less perfect wavy interface is formed it becomes clear its significant difference of its fine structure for any type of mutual solubility of the original metals. The corresponding images of the relief are shown in Fig. 4a for the tantalum interface of copper-tantalum joints and in Figs. 4b-d for the titanium interface of copper-titanium joints. If there is no mutual solubility, a wave does not contain any cusps inside it, and if there is mutual solubility, it does. This is the key role of generation of intermetallic compounds when the wavy interface is formed. Emergence of intermittent waves is another result of this process (Fig. 4d).

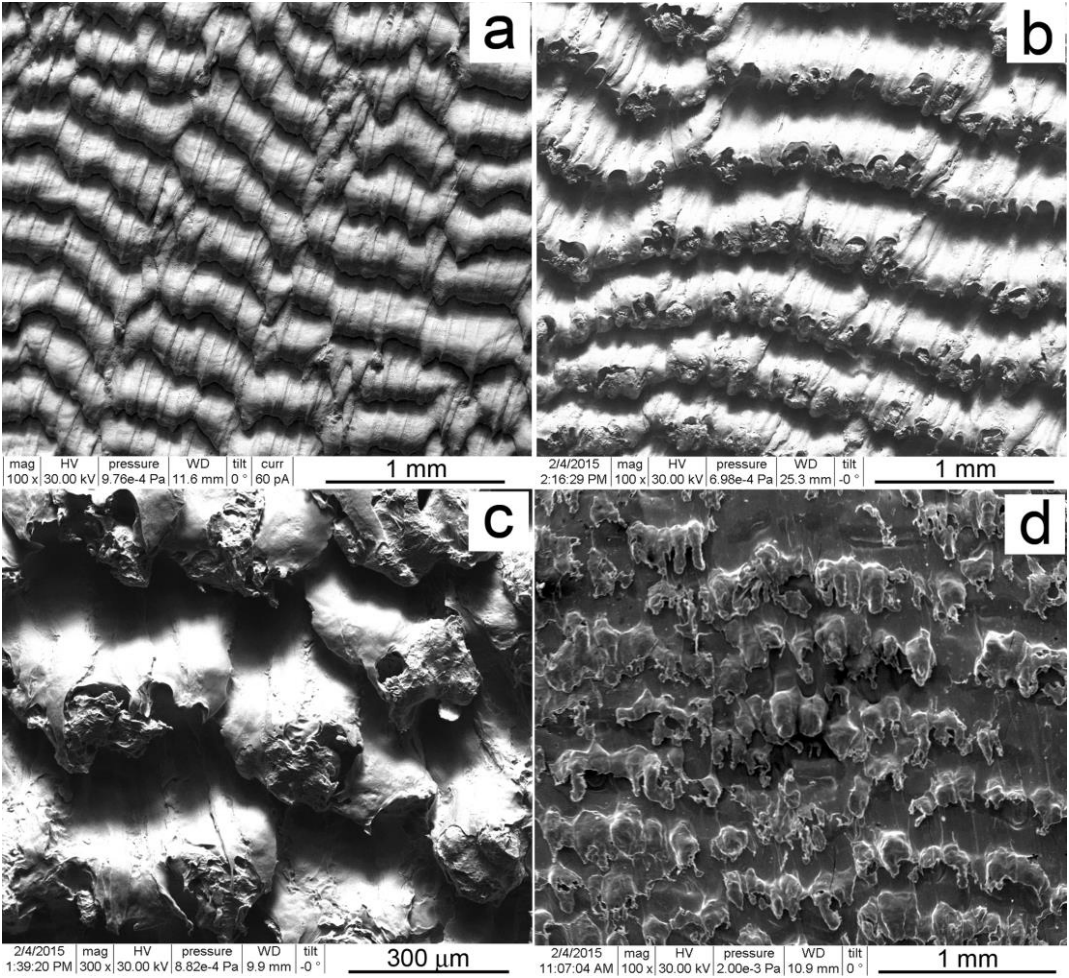

**Fig. 4.** Wavy interface: (a) - for copper-tantalum joints; (b) - for copper-titanium joints, cusps inside waves

Returning to images of the quasi-wave surface (Fig. 3), it can be noted an unexpected similarity to the surface relief in the aforementioned Abrahamson's experiments [4]. End views of steel bullet are given in Figs. 5a, b after a collision with a lead target under different angles of impact. Fig. 5c shows the steel slab surface after an impact of a thin copper sheet driven by means of explosion. To identify a relief observed in [4], a term "corrugations" is used. This relief resembles the "patchwork quilt" with alternating bands (Figs. 3c, d).

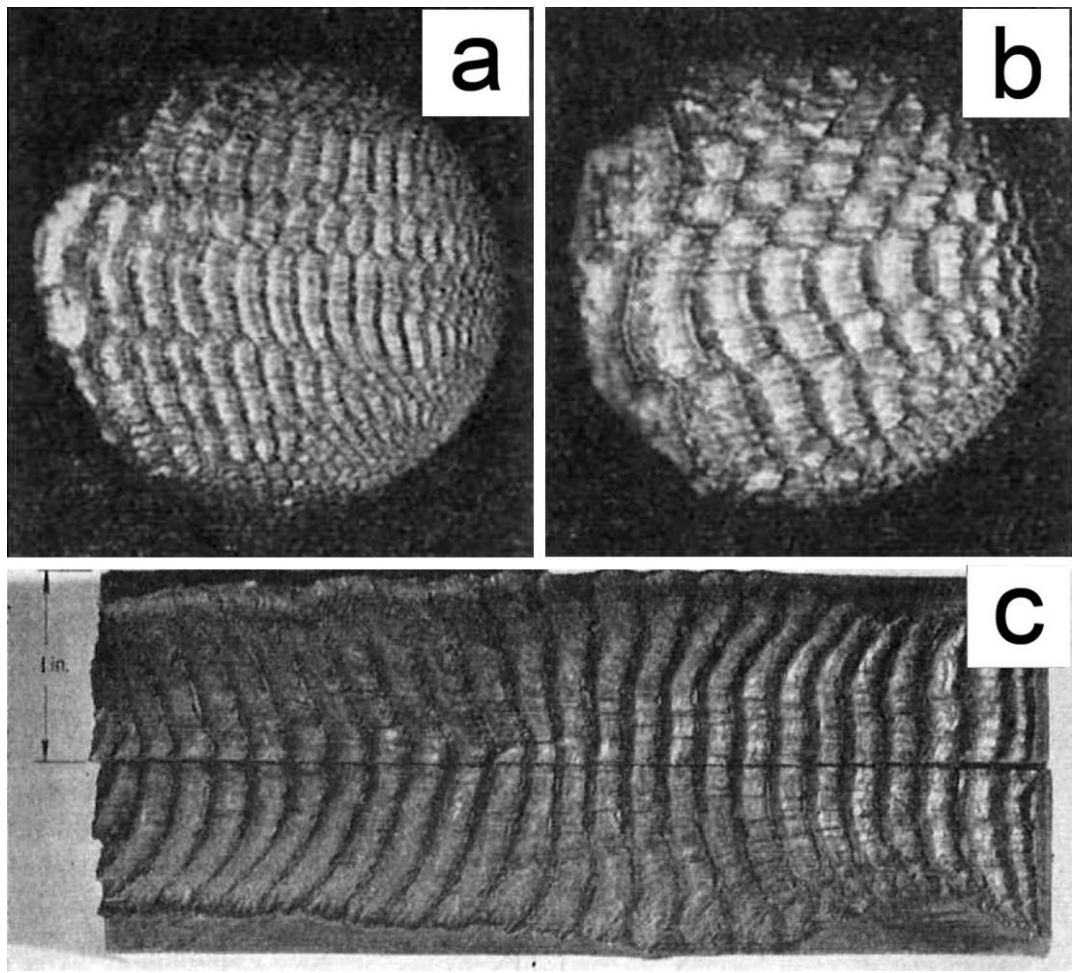

**Fig. 5.** Wavy interface of steel as a result of strong external impact: (a), (b) - colliding of steel bullet with lead plate (various angles of impact); (c) - impact of copper flying plate and steel slab

*3.2. Self-organization processes: transitions to waves*

It is seen from the electron-microscopic data that the evolution of the interface relief progresses with the following self-organization processes: from the cusps to the flashes, further to the big groups of flashes contacting each other, then to the quasy-wave interface, and finally to the more or less fairly perfect wavy interface. Before proceeding to a comparison of various transition states let us find out the reasons why they have not been observed previously. First of all, it is due to the fact that usually only the wavy interface is investigated because of its more practical application while the plain interface is not studied. As a result the welding parameters near the LB are not utilized. But in this case one should not expect observing either splashes or occasional cusps ensuring bonding. In order to examine these structures it should be possible to etch out one material entirely to observe the surface of the other. This may be achieved due to the high corrosion resistance of tantalum and titanium.

Besides, the wavy interface usually occurs at the modes within the weldability window. However, such an intermediate structure as the "patchwork quilt" is lost due to the fact that normally, the welding parameters in the range slightly higher than the LB are not used. The fact that the copper-tantalum joint features a relatively perfect wavy interface (Fig. 4a), different from the quasi-wavy interface, is one of the reasons of the stability of walls of the chemical reactor in heavy duty modes of operation [20].

It is worth noting that for different joints, even for those that possess a different type of the mutual solubility of the initial elements, the ways of forming the waves are quite similar. However, the different character of the mutual solubility of the elements at temperatures that occur in the contact area during the explosive welding, dramatically affects the structure of the waves formed. In the absence of mutual solubility of the elements at such temperatures, for example, for copper-tantalum joints, cusps appear first, followed by cusps and waves, and then - waves only. Figuratively speaking, one may say that a mechanism similar to that of a zip–fastener, works. Another situation occurs in case where the mutual solubility of the elements during the explosive impact is sufficiently high, and moreover, the intermetallics may arise in this system. An example of such a system is a copper-titanium joint, in which the cusps do not disappear during the formation of the waves. This is clearly seen in Figs. 4b-d. Figs. 4b, c shows a wave which is a dense pack of cusps pressed to each other, with their tips indented. The surface of the cusps is covered with fine, most likely intermetallic, particles. They prevent the action of a zip-fastener mechanism in the same way as dust does for real zip-fasteners, for instance. At a smaller external impact, another transition state is observed for another copper - titanium joint and this state can be called an intermittent wave (Fig. 4d). The wave is formed by cusps at a certain length, then breaks off, then cusps are formed again, etc.

It is the presence of the intermetallic particles that provides possible wave breaks, and thus providing for the existence of the intermittent waves. Thus, in the absence of mutual solubility of the elements the wave does not contain cusps within itself, while at the substantial solubility and abundance of the intermetallics in the system it does. This is the major role that the formation of intermetallic compounds plays in the creation of a wavy interface.

Let us demonstrate now the reasons why, during the explosive welding, and regardless of the nature of the mutual solubility of the materials, both cusps and splashes occur first, followed by the quasi-wave structure. The simplest assumption is that all of these structural features appear on the contact surface of the materials either in the process of the explosive impact or due to the relaxation of the occurring stress as well as due to the formation of the nonequilibrium structure. However, the mechanism, causing such a change of the form of the surface in a solid state, remains unclear.

First of all we would like to note that all of the observed structural changes indicate the occurrence of an excessive area of the interface in comparison with the original plain interface. It seems obvious that the explosive welding shock impacts on the contact surface are much higher than the yield stress of the materials. Therefore the possible increase of the interface, for example, as a result of deflection, will be irreversible. After the explosion load drops, a certain type of relaxation of the non–equilibrium structure will follow, while the surface area will remain excessive for reducing its energy. At the same time, the shape of the above surface must be changing, which may manifest itself by formation of cusps and splashes on the surface as well as by emerging of a wavy structure of the interface. Note that the solid phase transformations are supposed. It can be assumed that one of the reasons of occurrence of the said non-equilibrium surface structure is a rather large deflection of plates during their collision and the accompanying plastic deformation, localized near the contact surface.

### 3.3. *Imitating experiments*

Set of simulations have been conducted to determine possible ways of relaxation of a nonequilibrium structure with an excess area. Residual effects, occurring after some bending deformation of two contacting materials have been investigated.

A simple experiment has been performed: a metal plate was attached to a ruler at two fixed points. The thickness of the ruler was one order of magnitude more than that of the plate. A steel or plastic ruler was used. The plate was made of aluminum, copper, tantalum or lead. In any way, the ruler was much more flexible than the studied plate. The ruler was bent quit strongly and then released. As a result the ruler experienced only an elastic deformation and reverted to its original state after the stress was removed. At that, the less flexible plate experienced the plastic deformation and retained the excess area after the removal of the stress.

In those cases when the plate was not glued, all of it except the fixed points went through deflection which remained after the removal of the stress. As a result, the plate assumed a shape of a dome (Fig. 6a). The form of the plate that was initially glued to the ruler was completely different. Figs. 6-8 shows that the excess area is spent on the formation of several roof-like bends. It should be emphasized that the studied relaxation, consisting of formation of the bends, occurs at ambient temperatures and occupies a long period of time.

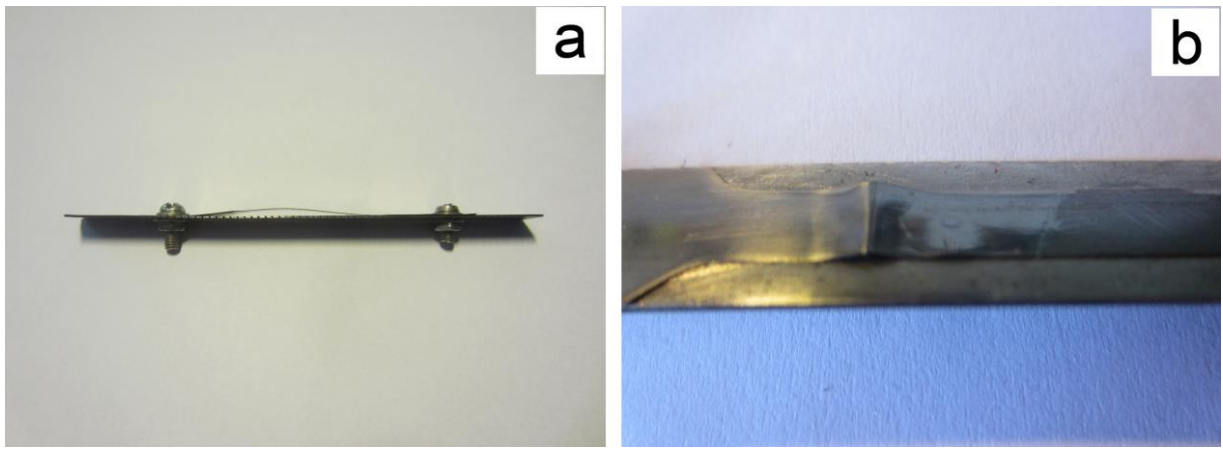

**Fig. 6.** Steel ruler - tantalum plate: (a) - deflection of a dome-like plate (plate is not glued); (b) - deflection of a roof-like plate (plate is glued)

Let us describe, at greater detail, the results of the simulation experiment for the pair that was made of the steel ruler and the tantalum plate. The ruler restored its flat form after deflection, while the plate that was not glued to the ruler, would assume the shape of a dome (Fig. 6a). If the plate was glued to the ruler the only bend occurred, and it had a shape of a "roof" (Fig. 6b). A video has been made for this pair which can be found at [21]. The sequence of frames corresponds to the scenario: the ruler, together with the tantalum plate glued to it, gets deflected, and then is gradually released, and at a certain moment a roof-type bend occurs on the plate.

Fig. 7 shows the results of the simulations for the "steel ruler - glued plate" pairs made of different metals: aluminum, copper, lead. In all cases the bends have been observed. Further, the experiment was extended to determine the difference in the behavior of the plate in case where it is glued to either the tensile ruler side, as in the previous cases, or to the compressed one. The ruler was made of plastic while the plate was made of lead. The thickness of the ruler was 2 mm and that of the plate was 0.1 mm. The radius of curvature was 30 mm for samples denoted by letter **A** and 20 mm for samples denoted by letter **B**. Figs.8a, b show the deflection of the plate for samples **A** and **B** respectively in the case when there is no gluing. The "dome" mentioned above is observed for sample **A**. For sample **B** the deflection has a more complex form, possibly due to the greater magnitude of the excessive length.

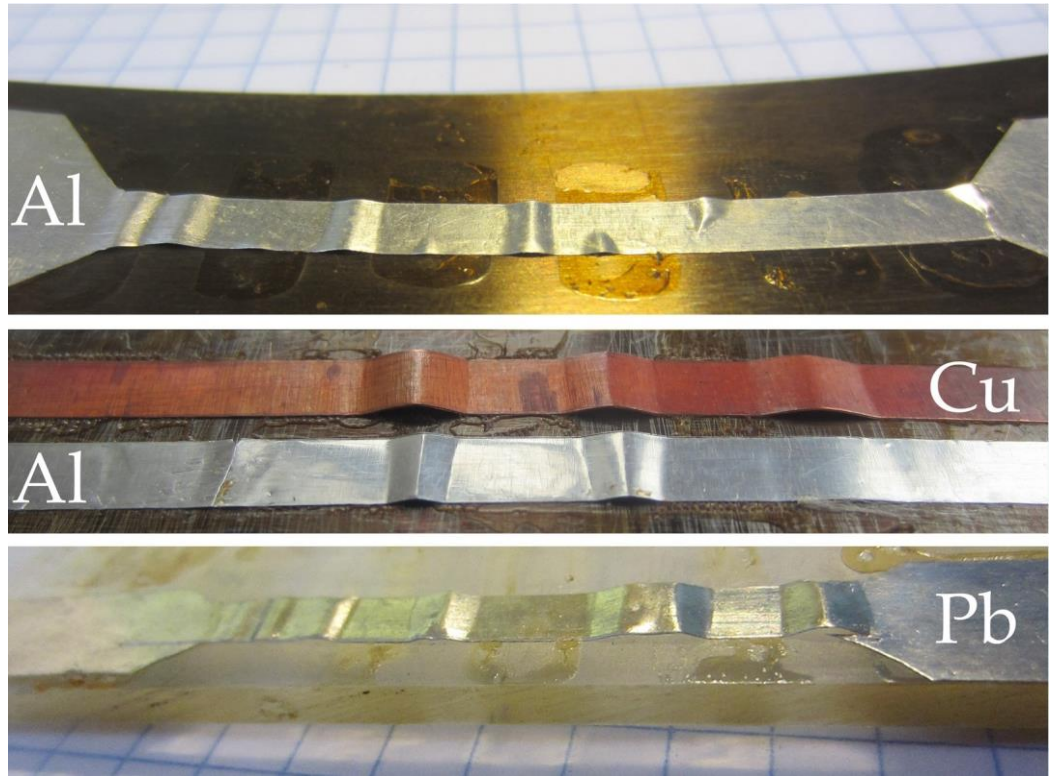

**Fig. 7.** Deflection of Al, Cu, Pb plates that are glued to steel ruler

As seen in Figs.8c-g the bends occur in cases where the plate is glued to the the tensile side of the ruler. If the plate is glued to the compressible side of the ruler, it still does not remain flat - its surface becomes wrinkled. Figs.8c-f show a shape of sample A plate that is glued both to the stretchable (Fig.8c) and the compressible (Fig.8e) sides of the ruler. Fig.8d shows a shape of the plate glued to the so-called midline [22]. As expected, in this case the plate remains flat, because there are no stresses at the midline. For the sample with a radius of curvature less than that of sample B we see the greatest number of bends, as seen on Fig. 8f. On the inner side of the bend (Fig. 8g) the relief changes only slightly.

Figs.6-8 show that the "roof" is hollow i.e., the bends occur due to a disjointing of the tested plate from the bearing ruler. This is due to the fact that the contact interface of the ruler and the plate is not uniform and both the formation of bend and disjoining occur in the weakest areas.

Simulation experiments, the results of which are presented here, at first glance, are not directly related to the explosive welding. However, their heuristic significance is in that they show a possibility to realise the excessive area by means of forming inhomogeneities on the contact surface.

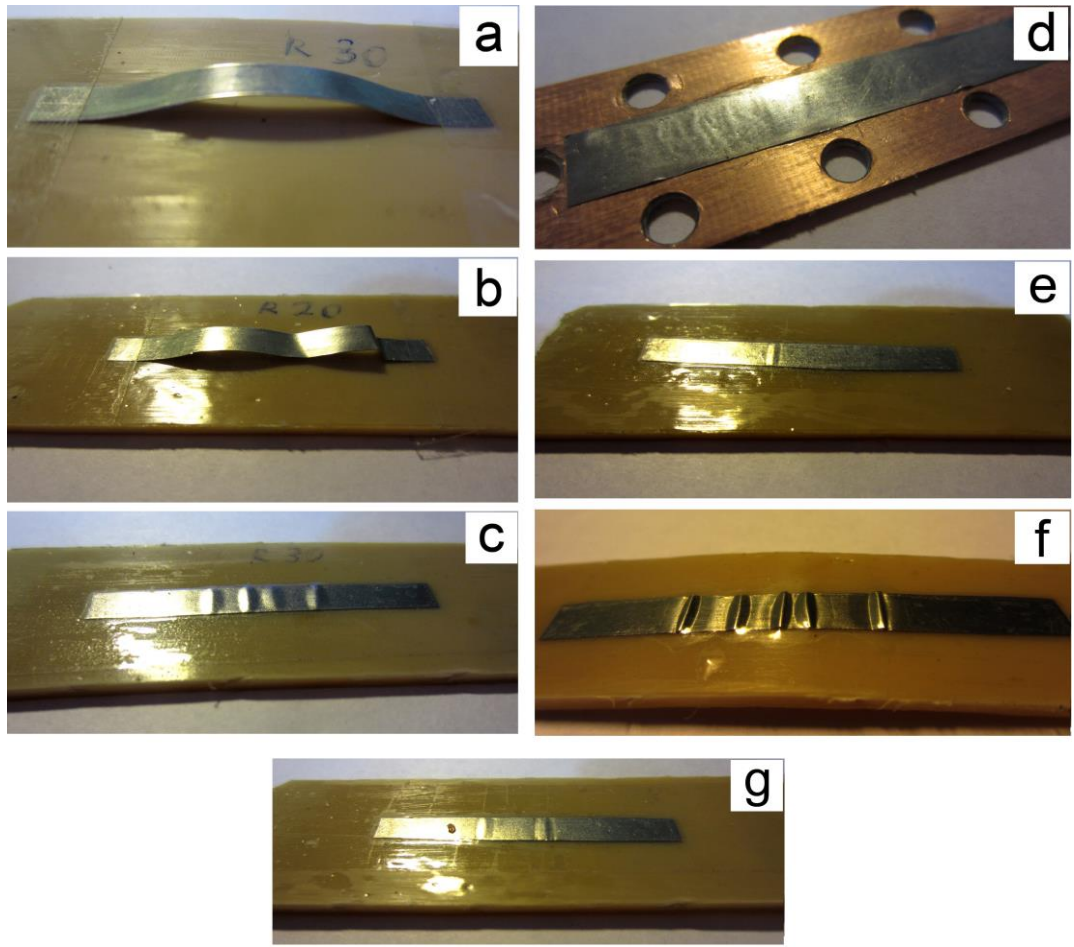

**Fig. 8.** Plastic ruler - lead plate: (a), (b) - deflection of plate (plate is not glued); (c) – (f) - deflection of plate that is glued to stretchable or compressible sides of the ruler, with a different radii of curvature

## 4. Conclusions

1. It was shown that the transition from the area below the LB to the area close to it is accompanied with a dramatic change of the interface relief: separate disconnected cusps after such a relatively low intensification of the welding mode are replaced by regularly distributed splashes. Being in a solid state they are really look like splashes on a water.

2. A very heterogeneous quasi–wavy surface is observed near the LB and slightly above it: waves have a different length and amplitude in different areas. Splashes can also be observed in some areas. The quasi-wavy interface in its structure resembles a "patchwork quilt" whose form depends on the type of the mutual solubility of initial elements.

3. It has been found that the quasi-wave interface of a "patchwork quilt" type observed in this work after explosive welding is similar to that of the steel bullet interface relief which was observed after collision with a target, at the famous Abrahamson's experiments [4].

4. Simulations have been conducted to determine possible ways of relaxation of a nonequilibrium structure having an excess area. Residual effects after a bending deformation of two contacting materials has been studied. It has been found that under a certain conditions the roof-type breaks occur on the plate of different metals (lead, aluminum, copper, tantalum). They imitate the heterogeneities on the interface arising upon the explosive welding.

**Acknowledgements**

Electron microscopic studies were performed at the Centre for Collective Use of Electron Microscopy of the Ural Division of the Russian Academy of Sciences and also at the Volgograd State Technical University.

The authors are grateful to the Russian Science Foundation (Project No.14-29-00158) for supporting the studies.